\documentclass[final]{svjour2}
\usepackage{graphicx}
\usepackage{rotating}
\usepackage{amssymb}
\usepackage{mathptmx}
\usepackage[numbers]{natbib}
\makeatletter
\journalname{Journal of Low Temperature Physics}

\bibpunct{}{}{,}{s}{}{,}

\begin{document}

\newcommand{\hdblarrow}{H\makebox[0.9ex][l]{$\downdownarrows$}-}
\title{Niobium Silicon alloys for Kinetic Inductance Detectors}

\author{M. Calvo$^1$ \and A. D'Addabbo$^{1,2}$ \and A. Monfardini$^1$ \and A. Benoit$^1$ \and N. Boudou$^1$ \and O. Bourrion$^3$ \and A. Catalano$^3$ \and L. Dumoulin$^4$ \and J. Goupy$^1$ \and H. Le Sueur$^4$ \and S. Marnieros$^4$}

\institute{1:Institut N\'eel \& Universit\'e Joseph Fourier, CNRS, BP 166, 38042 Grenoble, France\\
\email{martino.calvo@grenoble.cnrs.fr}
\\2: Universit\`a La Sapienza, Roma, Italy
\\ 3: Laboratoire de Physique Subatomique et de Cosmologie, Grenoble, France
\\ 4: Centre de Spectrom\'etrie Nucl\'eaire et de Spectrom\'etrie de Masse, Paris, France}
\maketitle

\begin{abstract}

We are studying the properties of Niobium Silicon amorphous alloys as a candidate material for the fabrication of highly sensitive Kinetic Inductance Detectors (KID), optimized for very low optical loads. As in the case of other composite materials, the NbSi properties can be changed by varying the relative amounts of its components. Using a NbSi film with T$_c \approx$ 1 K we have been able to obtain the first NbSi resonators, observe an optical response and acquire a spectrum in the band 50 to 300 GHz. The data taken show that this material has very high kinetic inductance $L_k$ and normal state surface resistivity $\rho_n$. These properties are ideal for the development of KID. More measurements are planned to further characterize the NbSi alloy and fully investigate its potential.


\end{abstract}

\section{Introduction}
First proposed in 2003 by the Caltech-JPL group\cite{Day03}, Kinetic Inductance Detectors (KID) have developed rapidly, reaching in the recent years a maturity level that already makes them ideal candidates for ground based astronomy. This has been shown for example by the NIKA experiment operating at millimetre wavelengths\cite{Monfardini13}. The NIKA detectors have been designed and optimized for working under high optical load. The background emission due to the atmosphere at the IRAM 30-m telescope site (2850 meters a.s.l. in the Sierra Nevada, Spain) is about 5 to 10 $\mathrm{pW}$ per pixel at 150 GHz. The NIKA detectors are based on Lumped Element resonators\cite{Doyle08}, whose geometry has been modified in order to make them dual-polarization sensitive\cite{Roesch12}. A thin (18nm) Aluminum film has been used to increase the value of $L_k$, and a strong coupling to the CPW feedline has been chosen to cope with the expected variations in the optical load, due mainly to variations in the atmospheric emissivity.

Building on this experience, we have started considering the possibility of using KID for space based missions. In space, the background power on the detectors and the signal that we expect to measure are up to two orders of magnitude lower than the ones observed on ground. The pixel design and fabrication therefore need to be thoroughly revised, starting from the choice of the materials used. Titanium Nitride has been first deployed in 2010 to make superconducting resonators\cite{Leduc10}. The physical properties of $Ti_{1-x}N_x$, such as its critical temperature $T_c$, can be changed by appropriately choosing the amount $x$ of nitrogen present. $Ti_{1-x}N_x$ has shown a surprisingly high value of kinetic inductance, $L_k\simeq 20 pH/\square$, and of normal state surface resistivity, $\rho_n \simeq 100\mu \Omega\, \mathrm{cm}$\cite{Leduc10}. The consequently high kinetic inductance fraction $\alpha = L_k/L_{tot}$ increases the responsivity of the detectors, making lower values of Noise Equivalent Power ($NEP$) achievable. In the case of LEKID, the larger $\rho_n$ represents a further advantage, as it makes directly coupling to the incoming radiation, in particular at shorter wavelengths, easier to achieve leading to increase optical efficiency. While undoubtedly a promising material, TiN has also some major drawbacks, in particular for what concerns the uniformity of the film composition and of its electromagnetic properties across wide areas, which are essential to fruitfully deploy the material for the fabrication of large arrays of detectors. Developments are ongoing, mainly in the US, in order to solve these problems. In this paper, we describe our first investigations for an alternative solution. For that, we take advantage of the long experience accumulated in France on NbSi films\cite{Marnieros98}.

\section{Device fabrication}

As in the case of $Ti_{1-x}N_x$, the properties of Niobium Silicon alloys ($Nb_xSi_{1-x}$) depend on the relative abundance of the components. This alloy exhibit a metal to insulator transition for $x\simeq 9\%$\cite{}. Above this value, the critical temperature grows linearly with $x$. Thus, $Nb_xSi_{1-x}$ can be used both for the fabrication of high impedance bolometers\cite{Bideaud11} and of Transition Edge Sensors (TES)\cite{nones}$^,$\cite{Martino13}. The films are deposited at CSNSM by co-evaporation of pure Nb and Si under ultra-high vacuum conditions. For our first test we chose a film thickness $t=50nm$, and a value of $x\simeq 18\%$. This corresponds to an expected $T_c$ of $\simeq 1K$, well matched to the base temperature of our dilution refrigerator and ideal for operation at millimetre wavelengths, starting from frequencies as low as $\simeq 70GHz$.

The design used for the first devices is a classical, single polarization sensitive LEKID geometry. The mask used had been actually designed for aluminum detectors. Some parameters of our device, such as the impedance of the CPW (CoPlanar Waveguide) feedline and the matching to the incoming radiation, were thus not optimized. Even though this leads to reflections on the line and poor optical efficiency, it does not represent any hindrance for establishing the fundamental properties of the compound under test.

\section{Electrical properties}

\begin{figure}[!t]
\begin{center}
\includegraphics[bb =73 55 555 751,
  width=0.4\linewidth, angle=-90,
  keepaspectratio]{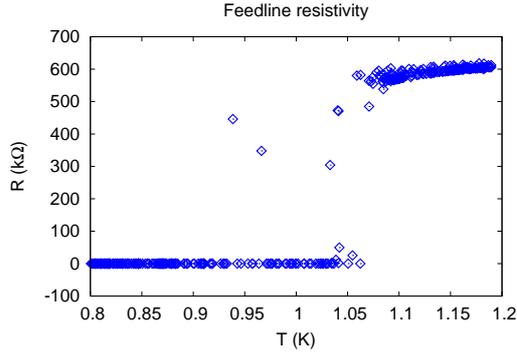}
\end{center}
\caption{(Color online) Resistivity of the $Nb_{.18}Si_{.82}$ as a function of temperature. The scatter of the data is due to the measurement system changing the readout current as a consequence of the rapid resistivity variation. The transition is evident at $T\simeq1.05K$. The residual normal resistance of the line just above the transition is $550 k\Omega$.}
\label{transiz}
\end{figure}

The first measurements have been carried out under dark conditions, with the chip mounted on the cold plate of a dilution refrigerator. The base temperature of this system is around $100\mathrm{mK}$. The resistance of a $RuO_2$ thermometer mounted on the cold plate is read using a four wire configuration. At the same time, and using the same configuration, we can monitor the resistivity of the CPW line used to excite the resonators. As shown in figure \ref{transiz}, the transition is observed at $T\simeq1.05K$, in very good agreement with our expectations. The transition is relatively sharp, i.e. less than 10 mK. The feedline has an overall length $l\simeq110 mm$ and a width $w=20 \mu \mathrm{m}$. Thus, we estimate the normal state resistivity of $Nb_{.18}Si_{.82}$ to be as high as $\rho_n\simeq 500 \mu \Omega \, \mathrm{cm}$.

\begin{figure}[!b]
\begin{center}
\begin{tabular}{cc}
\includegraphics[bb =2 265 545 579,
  width=0.48\linewidth,
  keepaspectratio]{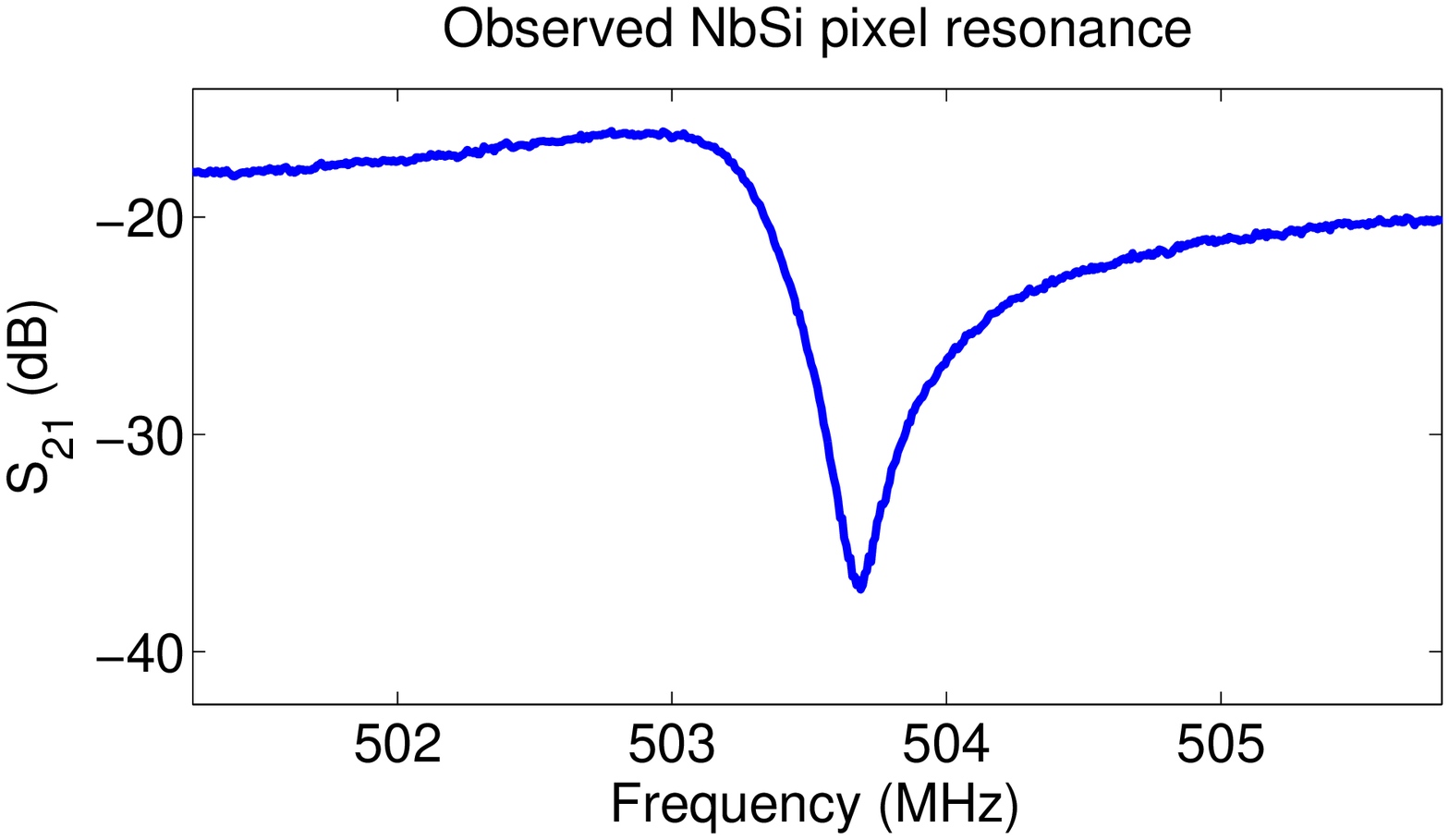}&
\includegraphics[bb =0 248 562 588,
  width=0.48\linewidth]{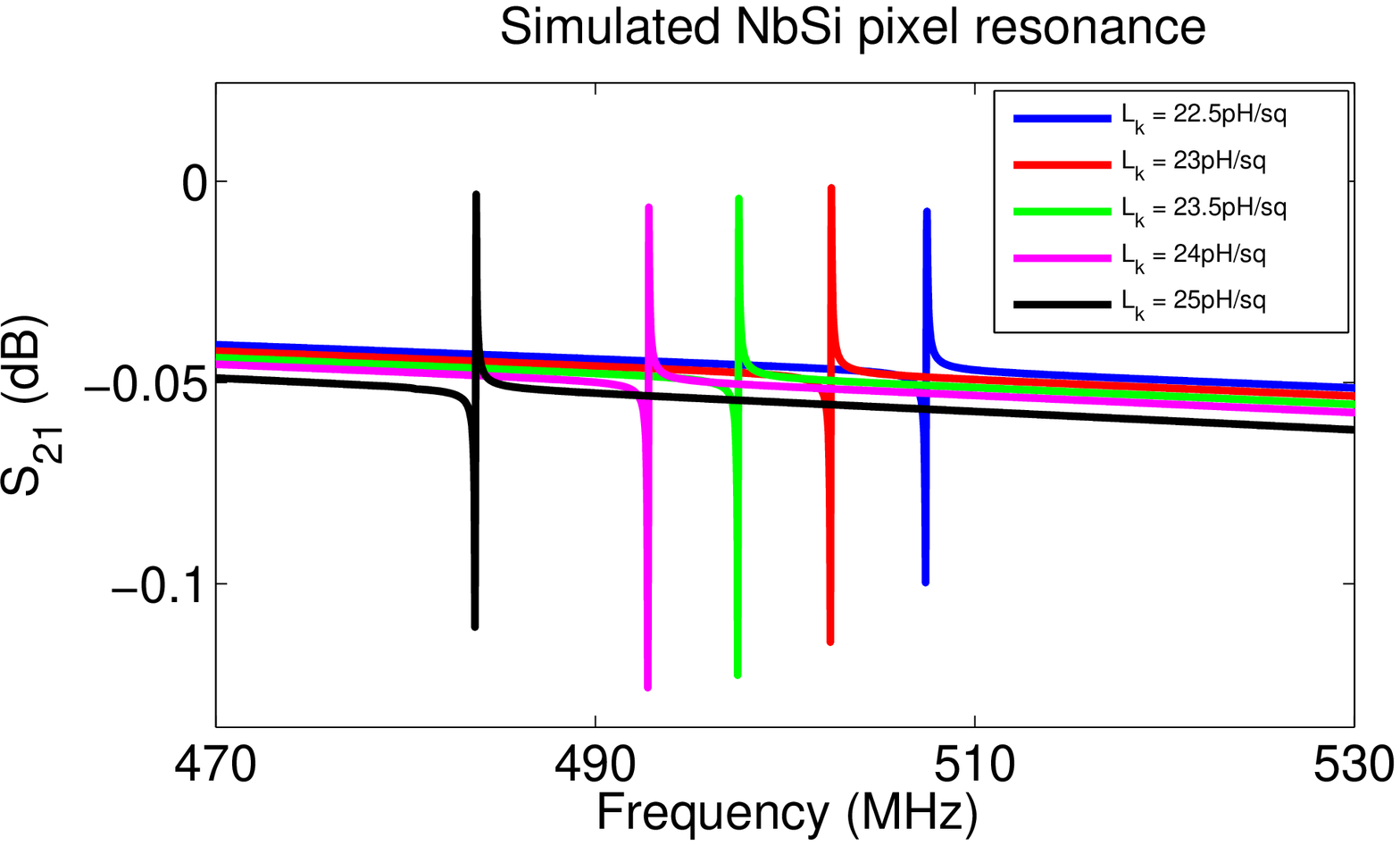}
  \end{tabular}
\end{center}
\caption{(Color online) Comparison between one of the observed resonances and the simulations carried  on a pixel having exactly the same geometry and $L_k$ values ranging between $22.5$ and $25\mathrm{pH}/sq$.}
\label{comparison}
\end{figure}

During a second cooldown, we connected the chip to the RF readout chain to be able to perform frequency sweeps with a Vector Network Analyzer (VNA). We can clearly identify the resonances, in a frequency interval lying between $500$ and $900\mathrm{MHz}$. For comparison, the very same resonator design lead, for $t=50nm$ Aluminum films, to the resonances lying in the $1.5\sim2\mathrm{GHz}$ range. This immediately indicates a large value of $L_k$. To actually estimate the kinetic inductance value we perform a series of simulations using the SONNET software, and vary $L_k$ in order to find the resonant frequencies near to the observed ones. The result is shown in figure \ref{comparison}, and it indicates that in our alloy $L_k\simeq 23\mathrm{pH}/sq$, which is comparable to the values obtained using $Ti_{1-x}N_x$. The shape of the resonances was strongly asymmetric because of the impedance mismatch due to the feedline, so that getting reliable estimates of the internal quality factor $Q_i$ was not possible.

\section{Optical properties}

\begin{figure}[!b]
\begin{center}
\includegraphics[bb =2 264 545 560,
  width=0.65\linewidth,
  keepaspectratio]{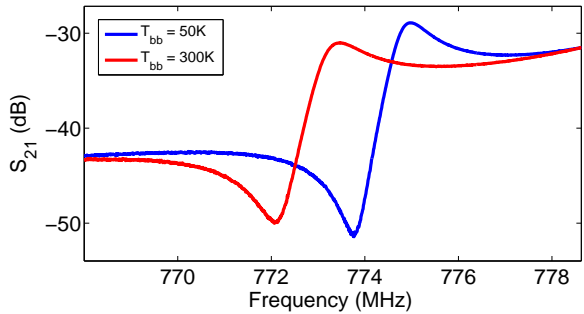}
\end{center}
\caption{(Color online) Effect of a change of optical load on the $Nb_{.18}Si_{.82}$ detectors. A frequency shift of $\Delta f = 1.7\mathrm{MHz}$ is measured when passing from a $300\mathrm{K}$ to a $50\mathrm{K}$ blackbody. The different values of $S_{21}$ below and above the resonances are a consequence of the off-resonance level not being flat. This is due to the standing waves introduced by the impedance mismatches in the readout line.}
\label{responsivity}
\end{figure}

We then mounted the $Nb_{.18}Si_{.82}$ chip in the cryostat designed for optical measurements, which is actually a copy of the one used for the NIKA camera at the IRAM 30m telescope. We measured the effect of observing two different blackbody sources, one at $300\mathrm{K}$ and one at $50\mathrm{K}$. The latter is obtained by means of a \textit{sky simulator}\cite{Monfardini11}, based on a dedicated Pulse Tube cooler. As can be seen in figure \ref{responsivity}, the frequency shift induced by the different optical loads is $\Delta f = 1.7\mathrm{MHz}$, which corresponds to a responsivity of $R \simeq 7\mathrm{kHz}/\mathrm{K}$. For comparison, using pixels of the same size but made of thin Al films ($12\sim 20 \mathrm{nm}$), and for the same variation of optical load, we obtained $R \simeq 1\mathrm{kHz}/\mathrm{K}$ at most.

\begin{figure}[!t]
\begin{center}
\includegraphics[bb =17 258 561 578,
  width=0.8\linewidth,
  keepaspectratio]{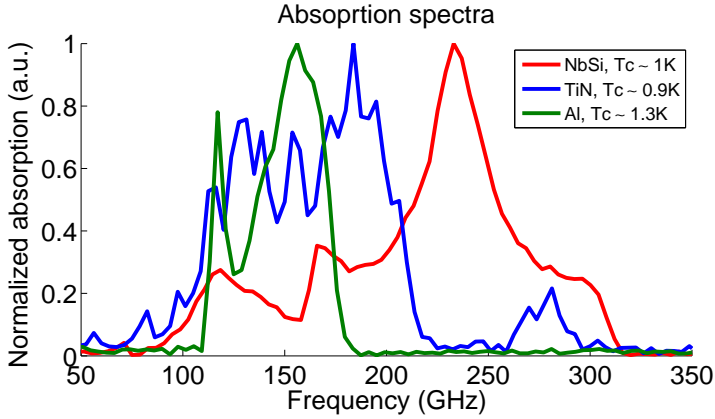}
\end{center}
\caption{(Color online) Absorption spectra for different materials. It is evident the different behaviour of compounds like $TiN$ and $NbSi$, whose absorption reduces smoothly below the gap frequency, from that of a pure metal like Al, which on the contrary shows a sharp cutoff at $\nu _{gap}$. The differences at higher frequencies are due to the different low-pass filters mounted in the various cooldowns and to the position of the $\lambda/4$ backshort that we mount to increase the absorption efficiency of our samples.}
\label{spectra}
\end{figure}

As a final step, we performed an absorption spectra to compare the properties of $Nb_{.18}Si_{.82}$ to those of other materials tested before. In this case a Martin-Pupplett Interferometer (MPI) is placed in front of the entrance lens of the cryostat. The MPI allows us to make spectra up to a few THz with a resolution which is typically set to $3\mathrm{GHz}$. The results show an interesting analogy with an effect that we had previously observed in TiN. As can be seen in figure \ref{spectra}, the behaviour of composite materials and that of a pure material like Al are remarkably different at frequencies below the cutoff value $\nu _{gap} = 3.5 k_b T_c/h$. In the case of Al we observe, as expected, a very sharp cutoff at $\nu _{gap}$, below which the material cannot absorb any incoming radiation as the photons are not energetic enough to break a Cooper Pair. This is in agreement with the classical description of a superconductor as a material with a well-defined energy gap dividing the superconducting carriers from the unpaired electrons. On the other hand, both for $Nb_{.18}Si_{.82}$ and for $TiN$, no clear cutoff is evident, although we still observe a clear suppression of absorption efficiency for frequencies well below $\nu _{gap}$. Possible explanations might be the presence of sub-gap states in compound materiales, or inhomogeneities in the deposited films leading to local variations of the critical temperature\cite{Sacepe08}. This point clearly needs further investigation to understand the underlying physical processes, before adopting such materials as the baseline for satellite missions.

\section{Future prospects}

Although the first results are very encouraging, more work needs to be carried out to confirm the potential of this alloy. In particular, the mask we used was not optimized for such a highly inductive material. This had two main consequences: first of all, the CPW feedline impedance was not $50\Omega$, so that strong reflections were present along the readout chain, leading to the onset of standing waves in the circuit affecting the uniformity of the detectors performances. And second, the resonances were shifted at frequencies well below $1\mathrm{GHz}$, out of the optimal operating range of the RF setup used. For this reason we have now designed a dedicated mask for this material, which has been optimized starting from the measured values of $\rho_n$ and $L_k$. With these new detectors we will be able to deepen our knowledge of the properties of $Nb_{x}Si_{1-x}$ alloys, and to obtain noise measurements and information about the quasiparticle lifetime, $\tau_{qp}$, and internal quality factor, $Q_i$.

\section{Conclusions}

We have tested the eletrical and optical properties of a new kind of alloys, $Nb_{x}Si_{1-x}$. This compound offers advantages similar to those of $Ti_{1-x}N_x$, in particular the possibility to tune $T_c$ changing $x$ and very high values of $L_k$ and $\rho_n$. We made samples starting from a $50\mathrm{nm}$ film of $Nb_{.18}Si_{.82}$, and measured $T_c\simeq 1.05\mathrm{K}$, $\rho_n\simeq 500 \mu \Omega \, \mathrm{cm}$ and $L_k\simeq 23\mathrm{pH}/sq$. The responsivity is also particularly high, $R \simeq 7\mathrm{kHz}/\mathrm{K}$, especially considering that the geometry was not optimized for matching the free space impedance.

$Nb_{x}Si_{1-x}$ has therefore many properties that make it ideal for Kinetic Inductance Detectors working at very low optical powers. In order to confirm the potential of this alloy, further measurements need to be carried out, in particular for what concerns its noise, the quasiparticle lifetime, and the $Q_i$ that can be achieved. To this end we have now designed a dedicated mask, optimized starting from the properties of the first film tested, which will allow us to better qualify this material and confirm its suitability for satellite based missions.


\begin{acknowledgements}
This work is been partially funded by the CNRS under an "Instrumentation aux limites 2013" contract, by the ANR under the contract NIKA and by the CNES R\&D program BSD. This work has been partially supported by the LabEx FOCUS ANR-11-LABX-0013.
\end{acknowledgements}


\end{document}